\documentclass[doublecol]{epl2}
\def\lesssim{\;\raise0.3ex\hbox{$<$\kern-0.75em\raise-1.1ex\hbox{$\sim$}}\;}
\def\gtrsim{\;\raise0.3ex\hbox{$>$\kern-0.75em\raise-1.1ex\hbox{$\sim$}}\;}

\revision

\title{Neutrino-pair bremsstrahlung in a neutron star crust}
\shorttitle{Neutrino-pair bremsstrahlung in neutron star crust} %Insert here a short version of the title if it exceeds 70 characters

\author{D. D. Ofengeim\inst{1,2} \and A. D. Kaminker\inst{1} \and D. G. Yakovlev\inst{1}}
\shortauthor{D. Ofengeim \etal}

\institute{
  \inst{1} Ioffe Institute,
Politechnicheskaya 26, 194021, St.\ Petersburg, Russia\\
  \inst{2} St.~Petersburg Academic University,  Khlopina Street 8/3,
St.~Petersburg 194021, Russia }


\abstract{Based on the formalism by Kaminker et al.\ (Astron.\
Astrophys.\ {\bf 343} (1999) 1009) we derive an analytic
approximation for neutrino-pair bremsstrahlung emissivity due to
scattering of electrons by atomic nuclei in the  neutron star crust
of any realistic composition. The emissivity is expressed through
generalized Coulomb logarithm which we fit by introducing an
effective potential of electron-nucleus scattering. In addition, we
study the conditions at which the neutrino bremsstrahlung in the
crust is affected by strong magnetic fields. The results can be
applied for modelling of many phenomena in neutron stars, such as
thermal relaxation in young isolated neutron stars and in accreting
neutron stars with overheated crust in soft X-ray transients.}

\begin{document}

\maketitle

%%%%%%%%%%%%%%%%%%%%%%%%%%%%%%%%%%%%%%%%%%%%%%
\section{Introduction}

It is well known that studies of the thermal evolution of neutron
stars allows one to explore the physics of their superdense matter
\cite{Haens2007}. The thermal evolition, in turn, is largely
determined by neutrino emission from various neutron star layers
\cite{Yak2001}. Here we focus on the most important neutrino
emission mechanism in the neutron star crust, which is the emission
of neutrino-antineutrino pairs (of any flavors, through charged and
neutral electroweak currents) in collisions of electrons (e) with
atomic nuclei ($A,Z$). It is also called the neutrino-pair
bremsstrahlung, written schematically as
\begin{equation}
\label{bremsScheme}
  {\rm e} + (A,Z) \rightarrow {\rm e} + (A,Z) + \nu +
\widetilde{\nu}.
\end{equation}
It was proposed by
Pontecorvo \cite{Pontecorvo1959} and has been studied in astrophysical context
many times as  reviewed in \cite{Yak2001}.

We employ the formalism of \cite{Kam1999} which
includes the most complete collection of physical effects
important in (\ref{bremsScheme}) and
allows one to calculate the neutrino bremsstrahlung
emissivity $Q$ [erg s$^{-1}$ cm$^{-3}$] for any composition of
neutron star crust. To use these results in
modeling of neutron star phenomena, one needs either extensive tables or
analytic fits of $Q$. The authors of \cite{Kam1999} fitted
 $Q$ for the ground-state crust. However, the composition may differ from the ground
state; for instance, the crust can be accreted \cite{Haens2007}.
Accordingly, one needs to fit $Q$ for a wide
range of possible compositions. In this Letter we
calculate $Q$ for different compositions
 and obtain the required
fit. In addition, we investigate the conditions at which
the process (\ref{bremsScheme}) is affected by
magnetic fields.

%%%%%%%%%%%%%%%%%%%%%%%%%%%%%%%%%%%%%%%%%%%%%%%%
\section{Formalism}
%%%%%%%%%%%%%%%%%%%%%%%%%%%%%%%%%%%%%%%%%%%%%%%%

We analyze the process (\ref{bremsScheme}) in the crust of a neutron
star without magnetic field (but we discuss the magnetic effects in
the end of the Letter).  Under the crust we mean the envelope of the
star which contains atomic nuclei. It extends \cite{Haens2007} to
the density $\rho \approx 1.5 \times 10^{14}$~g~cm$^{-3}$, about
half of the saturated nuclear matter density. The atomic nuclei
there are crystallized or form Coulomb liquid.

At any value of $\rho$ the matter is assumed to contain  spherical
atomic nuclei of one species.  The nuclei are immersed in the sea of
electrons, and at densities higher than the neutron drip density
$\rho_{\rm ND} \approx (4-6)\times 10^{11}$ g~cm$^{-3}$
\cite{Yak2001}, also in the sea of free neutrons.

Following \cite{Kam1999} we will restrict ourselves to the case
of ultrarelativistic  and
strongly degenerate electrons. In this case
 $p_{\rm F}\gg m_{\rm e}c$, that is  $\rho \gg 10^6$
g~cm$^{-3}$, where $p_{\rm F}=\hbar (3 \pi^2 n_{\rm e})^{1/3}$ is
the electron Fermi momentum, $m_{\rm e}$ the electron rest-mass,
and $n_{\rm e}$ is the electron number
density. Such electrons are strongly degenerate at
$cp_{\rm F}\gg k_{\rm B}T$, $T$ being the temperature and $k_{\rm
B}$ the Boltzmann constant. Under these conditions the electrons
form nearly ideal Fermi gas and the nuclei (ions) are fully ionized
by the electron pressure. The electric neutrality of the matter
implies $n_{\rm e}=Zn_{\rm i}$, where $n_{\rm i}$ is the number
density of the nuclei, and $Z$ is the nucleus charge number. Let us
introduce also $A_{\rm tot}$, the total number of nucleons per one
nucleus, and $A=A_{\rm nuc}$, the total number of nucleons confined in one
nucleus. In the outer crust ($\rho < \rho_{\rm ND}$) we have $A_{\rm
tot}=A$, while in the inner crust ($\rho \ge \rho_{\rm ND}$) one
gets $A_{\rm tot}>A$ (with $A_{\rm tot}-A$ being the number of free
neutrons per one nucleus). The mass density of the matter in the
crust is $\rho \approx m_{\rm u} A_{\rm tot} n_{\rm i}$,
where $m_{\rm u}$ is the atomic mass unit.

Let us introduce the dimensionless parameters
\begin{eqnarray}
  x & = & \frac{p_{\rm F}}{m_e c}  = 25.73 \left( Z n_{\rm i\,34} \right)^{1/3},
\label{xDef} \\
  \Gamma & = & \frac{Z^2 e^2}{k_{\rm B}Ta}
    =0.5798\, \frac{Z^2}{T_9} \, n_{\rm i\,34}^{1/3},
 % = 2.275\, \frac{Z^2}{T_9} \left( \frac{Z\rho_{12}}{A_{\rm tot}} \right )^{1/3},
  \label{GDef} \\
  \tau & = & \frac{T}{T_{\rm pi}} =
      0.9914\,\frac{T_9}{Z} \, \sqrt{\frac{A}{n_{\rm i\,34}}},
 % 1.277\,\frac{T_9}{Z}\sqrt{\frac{A_{\rm tot}A}{\rho_{12}}},
\label{tDef} \\
  \xi & = & \frac{R_{\rm p} p_{\rm F}}{\hbar}
  = \left( \frac{9\pi}{4} \,Z \right)^{1/3} \frac{R_{\rm p}}{a}
\nonumber \\
%   =   0.261 {R_{\rm p ~fm}}\, \left( \frac{Z \rho_{12}}{A_{\rm tot}}  \right)^{1/3}.
   & & = 0.06665 R_{\rm p\;{fm}} \left( Z n_{\rm i\,34} \right)^{1/3}.
\label{xiDef}
\end{eqnarray}
Here $a=(4\pi n_{\rm i})^{-1/3}$ is the ion sphere (spherical
Wigner-Seitz cell) radius; $T_{\rm pi}=\hbar \omega_{\rm pi}/k_{\rm
B}$ is the ion plasma temperature determined by the ion plasma
frequency $\omega_{\rm pi}=\sqrt{4 \pi Z^2 e^2 n_{\rm i}/(A m_{\rm
u})}$; $n_{\rm i\,34}=n_{\rm i}/(10^{34}~{\rm cm}^{-3})$; $T_9 =
T/(10^9~{\rm K})$; $R_{\rm p\;{fm}}=R_{\rm p}/(1~{\rm fm})$ is the
proton core radius of the nucleus expressed in fm. Thus,
$x$ is the relativity parameter of degenerate electrons ($x \gg 1$
in our case); $\Gamma$ is the Coulomb coupling parameter ($\Gamma
\lesssim 1$ for ion gas; $\Gamma \gtrsim 1$ for ion liquid or
crystal; crystallization occurs at $\Gamma=175$); $\tau$ determines
the importance of quantum effects in ion motion; $\xi$ specifies the
importance of finite sizes of the nuclei; see \cite{Haens2007}
for details.

According to \cite{Kam1999}, the  emissivity of the process
(\ref{bremsScheme}) is
\begin{eqnarray}
 Q & = & \frac{8\pi G_{\rm F}^2 e^4 C_{+}^2}{567 \hbar^9 c^8}\, Z^2
 (k_{\rm B}T)^6 n_{\rm i}\, L(Z,A,R_{\rm p},n_{\rm i},T)\,R_\mathrm{NB}
\nonumber
\\
  & = & 5.362\times 10^{15} Z^2 n_{\rm i\,34} T_9^6 L R_{\mathrm{NB}} \;
    %3.299 \times 10^{17} Z^2 \rho_{12} T_9^6 L R_{\mathrm{NB}}/A_{\rm tot} \;
  {\mbox{erg}}\,{\mbox{cm}^{-3}\mbox{ s}^{-1}}.
\label{Qmain}
\end{eqnarray}
Here $G_{\rm F}$ is the Fermi weak interaction constant; $C_+^2 =
1.675$ takes into account three neutrino flavors; $R_{\mathrm{NB}} =
1+0.00554Z+0.0000737Z^2$ is an approximate non-Born correction.
Furthermore, $L$
is the generalized Coulomb logarithm to be determined; $L = L_{\rm
liq}$ in the liquid phase; $L = L_{\rm ph}+L_{\rm sl}$ in the solid
phase, where $L_{\rm ph}$ is the contribution of electron-phonon
scattering, and $L_{\rm sl}$ is the contribution of Bragg diffraction
of electrons on the static lattice. In the liquid phase,
\begin{equation}
\label{Lliq}
  L_{\rm liq} = \int_0^1 {S_\mathrm{liq}(q)} |V(q)|^2
  {\cal R}_{\rm c}(y) y^3\,{\rm d}y,
\end{equation}
where $\hbar q$ is the electron momentum transfer in a reaction
event and $y = \hbar q /(2 p_{\rm F})$.
The function $S_\mathrm{liq}(q)$ is the ion structure factor in the
Coulomb liquid \cite{Young1991}; it takes into account
ion-ion correlations which introduce strong ion screening
of the electron-nucleus interaction; ${\cal R}_{\rm c}(y) = 1 + 2y^2 (\ln
y)/(1-y^2)$ comes from the squared matrix element;
$V(q) = F(q)/(y^2+y_0^2)$ is the Fourier transform of
the Coulomb potential screened by electron polarization (included
into $y_0$ \cite{Kam1999});
$F(q)$ is the nuclear form factor which takes
into account proton charge distribution within the nucleus and
associated additional effective screening \cite{Kam1999}.

Let us consider the proton charge distribution within the nucleus as
uniform (uniformly charged sphere of radius $R_{\rm p}$). Then $F(q)
= F(u) = 3 (\sin u - u \cos u)/u^3$, $u = 2\xi y$. At $\rho \lesssim
10^{12}$ g~cm$^{-3}$ one typically has $R_{\rm p} \ll a$ and finite
sizes of atomic nuclei are unimportant. Then it is sufficient to set
$R_{\rm p} \to 0$ ($\xi \to 0$, $F(q)=1$), i.e., treat the nuclei as
point-like. At $\rho \gtrsim 10^{12}$ g~cm$^{-3}$, the nuclei occupy
a non-negligible volume which introduces extra effective screening
of the electron interaction with the nuclei. In the density range
$10^{12} \lesssim \rho \lesssim 10^{13}$ g~cm$^{-3}$ the
approximation of uniformly charged proton core is reasonably good
but at higher $\rho$ it breaks down (the proton cores are greatly
broadened by the density effects \cite{Haens2007}). However, in this
case one can take realistic proton charge distribution, calculate
its root-mean-square value, $\sqrt{\langle R_{\rm p}^2 \rangle}$,
and use this value instead of $R_{\rm p}$ in the formulas obtained
formally for the uniform charge distribution. This turns out to be a
good approximation \cite{Gned2001} which allows us to use the
employed formalism even if real nuclei are charged non-uniformly.

The expression for $L_{\rm ph}$ is
\begin{equation}
\label{Lph} L_{\rm ph} = \int_{(4Z)^{-1/3}}^1 S_{\rm ph}(q) |V(q)|^2
R_{\rm c}(y) y^3\, \mathrm{d}y.
\end{equation}
An appropriate effective structure factor $S_{\rm ph}$ is obtained
in \cite{Bai1998} and takes into account multiphonon processes,
\begin{equation}
\label{Sphon}
  S_{\rm ph} = \left[ \exp(w_1 y^2) - 1 \right] \exp(-w y^2),
\end{equation}
\begin{equation}
\label{w1phon} w_1 = \left(12\pi^2\right)^{1/3} \frac{Z^{2/3}u_{-2}}
{\Gamma} \frac{b \tau}{\sqrt{(b\tau)^2 + u_{-2}^2 \exp (-7.6\tau)}},
\end{equation}
\begin{equation}
\label{wphon} w = \left(12\pi^2\right)^{1/3} \frac{Z^{2/3}u_{-2}}
{\Gamma} \left( \frac{u_{-1}}{2u_{-2}}\frac{\exp(-9.1\tau)}{\tau} +
1 \right),
\end{equation}
where $b=231$, $u_{-1}=2.798$, and $u_{-2}=12.972$ ($u_{-1}$ and
$u_{-2}$ being dimensionless moments of phonon frequencies of the
Coulomb crystal). The crystal is assumed to have the body centered
cubic structure, but the results are fairly insensitive {\bf to} the
lattice type \cite{Kam1999}.

Finally,
\begin{equation}
\label{Lsl}
  L_{\rm sl} = \frac{1}{12Z}\sum_{\bm{K}\ne 0}
  (1-y^2)y^2 |V(K)|^2 I(t_V,y) \exp(-wy^2),
\end{equation}
where $\bm{K}$ is a reciprocal lattice vector, $y = \hbar
|\bm{K}|/(2p_{\rm F})$,
\begin{equation}
\label{tV} \frac{1}{t_V} = \frac{\Gamma}{Z} \left( \frac{4}{3\pi Z}
\right)^{2/3} \sqrt{1-y^2} \,|V(K)| \exp(-wy^2).
\end{equation}
The function $I(t_V,y)$ has been analyzed in
\cite{Kam1999}; it takes into account electron band structure effects
(Bloch states instead of plain waves). These effects
broaden Bragg diffraction peaks and reduce $L_{\rm sl}$ \cite{Pethick1997}.

Note that the approach  \cite{Kam1999} does not take into account
the quantum effects of ion motion in the liquid state [$S_{\rm
liq}(q)$ in (\ref{Lliq}) is classical, independent of $\tau$]. This
approximation may actually be violated, especially for light ions
near the melting point. However, for heavy ions the approximation is
justified at all $T$ in the liquid. Note also that in our case $L$
is explicitly independent  of $x$.

We will calculate and fit the Coulomb logarithm $L(Z,A,R_{\rm p},n_{\rm
i},T)$. It can be presented as a function of four
dimensionless arguments, $L=L(Z,\Gamma,\tau,\xi)$.

%%%%%%%%%%%%%%%%%%%%%%%%%%%%%%%%%%%%%%%%%%%%%%%%%%%%%%%%%%%%%
\section{Coulomb logarithm and effective potential}
%%%%%%%%%%%%%%%%%%%%%%%%%%%%%%%%%%%%%%%%%%%%%%%%%%%%%%%%%%%%%%%

Our fit of the Coulomb logarithm will use the
concept of effective potential for the electron-nucleus scattering
in the process (\ref{bremsScheme}). Similar effective potentials
have been introduced by A.~Y. Potekhin to fit
Coulomb logarithms \cite{Gned2001} for
electric and thermal conductivities of electrons due to electron-ion
(electron-phonon) scattering in the crust. Although the
principal approach will be the same, the effective potentials are
naturally different.

In our case $V_{\rm eff}$ is introduced as
\begin{equation}
\label{Lappr}
 L = \int_0^1 \left| V_{\rm eff}(y) \right|^2 {\cal R}_{\rm c}(y)
 y^3\, \mathrm{d}y,~~~
  |V_{\rm eff}|^2 = S_{\rm eff}(y) \frac{|F(2\xi y)|^2}{y^4},
\end{equation}
where $S_{\rm eff}$ is the effective structure factor which we
present in the form close to (\ref{Sphon}),
\begin{equation}
\label{Seff} S_{\rm eff} = \left[ \exp({w}_1^* y^2) - 1 \right]
\exp(-{w}^* y^2).
\end{equation}
Then  ${w}^*_1$ and ${w}^*$ can be taken
similar to (\ref{w1phon}) and (\ref{wphon}),
\begin{equation}
\label{w1eff}
 {w}_1^* = B \; \frac{b\tau_*}{\sqrt{ \left(b\tau_*\right)^2
 + u_{-2}^2 \exp\left[-7.6\left(\tau_*
 +{18}/{\Gamma_*}\right)\right] }},
\end{equation}
\begin{equation}
\label{weff}
  {w}^* = B\; \left[1+ \frac{u_{-1}}{2u_{-2}\tau_*}
  \exp\left(-9.1\left\{ \tau_*+\frac{216}{\Gamma_*}\right\} \right) \right],
\end{equation}
\begin{eqnarray}
  B &= & \frac{3}{5}\frac{( 12\pi^2 )^{1/3}u_{-2}Z^{4/5}}{1+({\xi}/{Z})
  {\Gamma_*}/(200 + \Gamma_*)}
\nonumber \\
  & \times &
  \left( \Gamma_*^2 +
  \frac{1037}{\left( 1+{\Gamma_*}/{204} \right)^4}\,
  \frac{\Gamma_*^{(1+0.15\xi)/2}}{Z^{0.10}} \right)^{-1/2}.
\label{BforW}
\end{eqnarray}
Here $\tau_*$ and
$\Gamma_*$ are rescaled $\tau$ and $\Gamma$, respectively.
They are related to original $\Gamma$, $\tau$, $\xi$ and $Z$ as
\begin{equation}
\label{tEff}
  \tau_* = 0.095 \left( \frac{2\tau}{0.095\sqrt{\tau^2 + 4}}
  \right)^{1/\Lambda}, \quad \Lambda=1+\frac{(Z/35.5)^2}
  {1+{\Gamma}/({223Z})},
\end{equation}
\begin{equation}
\label{GEff}
   \Gamma_* = \Gamma
   \exp\left\{ - \frac{\tau_*^{0.053}
   \left({Z}/{11.3}\right)^{4/9}}
   {{1+\tau_*}^G \left[\Gamma/(19.3Z^{1.7})\right]^H} \right\},
\end{equation}
with $G=0.5+0.002Z\xi$ and $H=[1.52+0.9(1-\tau_*)]/(1+0.5\xi)$.

In the high-$\Gamma$ limit at $\xi=0$ equations (\ref{w1eff}) and (\ref{weff})
reduce to (\ref{w1phon}) and (\ref{wphon}), respectively, except that
$Z^{2/3}$ transforms to $(3/5) Z^{4/5}$.

For an analytic integration of the Coulomb logarithm (\ref{Lappr})
we suggest to approximate ${\cal R}_{\rm c}$ as  ${\cal R}_{\rm
c}(y)\approx 1-y$ (with a maximum relative error equal to $0.08$ at
$y=0.3$ over $0\leq y \leq 1$) and the form factor as
$|F(u)|^2\approx \exp(-\alpha u^2 )$ with $\alpha = 0.23$ (maximum
absolute error equal to $0.032$ at $u=1.6$ for any $u \geq 0$). Then
(\ref{Lappr}) takes the form
\begin{eqnarray}
L &= & f\left(w^*+4\alpha\xi^2\right) -
f\left(w^*-w^*_1+4\alpha\xi^2\right),
\label{LapprInt} \\
f(x) & = & \frac{1}{2} \left[ \sqrt{\frac{\pi}{x}} \;
\mathrm{erf}(\sqrt{x}) + \ln(x) - \mathrm{Ei}(-x) \right], \nonumber
\end{eqnarray}
where erf$(x)$ and Ei$(x)$ are the standard error function and
exponential integral, respectively. Note that $f(x)=0.71139+x/6$ as
$x \to 0$.

\begin{figure}
\onefigure[width=0.45\textwidth]{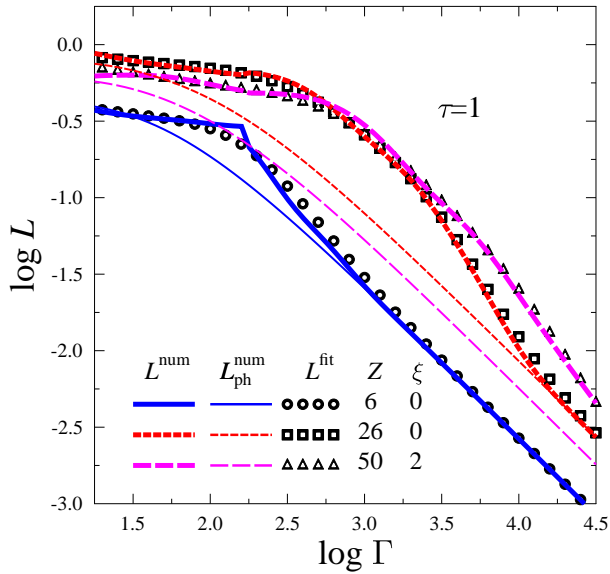}%{lgL-lgG_exact+phon}
\caption{(Color online) Coulomb logarithms versus $\Gamma$ in
log-log scale at $\tau=1$ for three types of nuclei. Shown are the
full logarithm $L^{\rm num}$ (thick lines) and its formal phonon
contribution $L^{\rm num}_{\rm ph}$ (thin lines) calculated
numerically using the formalism \cite{Kam1999}, as well as our
fits $L^{\rm fit}$ to the full logarithm
(symbols). The solid lines and open dots refer to the nuclei with $Z$=6
at $\xi=0$; the short-dashed lines and squares are for $Z=26$
and $\xi=0$, while the long-dashed lines and triangles are for $Z$=50 at $\xi=2$.
See text for details.}
\label{fig_logars}
\end{figure}

To clarify the proposed fit let us note that $L_{\rm ph}$
roughly reproduces the full Coulomb logarithm $L$ (fig.\
\ref{fig_logars}). Accordingly, $L_{\rm ph}$ is basic for our
fit. Modified $w^*$ and $w_1^*$ allow us to accurately reproduce
$L_{\rm liq}$, while $\Gamma_{*}$ and $\tau_{*}$
tune the fit of $L_{\rm sl}$. More details are given below.

The fit contains a number of coefficients which have been initially
chosen ``by eye'' but have been optimized afterwards by comparing
with calculations based on the formalism \cite{Kam1999}. The
comparison has been made on the following grid of the parameters:
$Z$=6,\ldots,50 (10 points), $A=1.9Z,\ldots, 3Z$ (10 points),
$R_{\rm p}=(0.0,\ldots,0.2)\times a$ (10 points), $\log n_{\rm
i}[{\rm cm}^{-3}] = 29,\dots,34$ (15 points), and $\log T[{\rm K}] =
8,\dots,9$ (15 points). Also, we have excluded the cases in which
$Q<10^7$ erg~cm$^{-3}$~s$^{-1}$  and/or $|\log \Gamma/175| < 0.2$.
The first condition excludes the range of parameters where the
neutrino bremsstrahlung is too slow (unimportant for applications).
The second condition excludes narrow temperature intervals near the
melting points. In these intervals the formalism  \cite{Kam1999}
gives relatively small jumps of $Q$ which  are believed to be
artificial, insignificant for applications and caused by neglecting
the quantum effects in ion motion in the liquid phase (see above).
Our exclusion condition smoothes out such jumps which seems adequate
to the physics of the problem. After these exclusions, the
optimization has been provided on 172550 grid points by comparing
the values of $L$, calculated via (\ref{Lliq})--(\ref{Lsl}), with
those given by the fits (\ref{w1eff})--(\ref{LapprInt}).

\begin{figure}
\onefigure[width=0.45\textwidth]{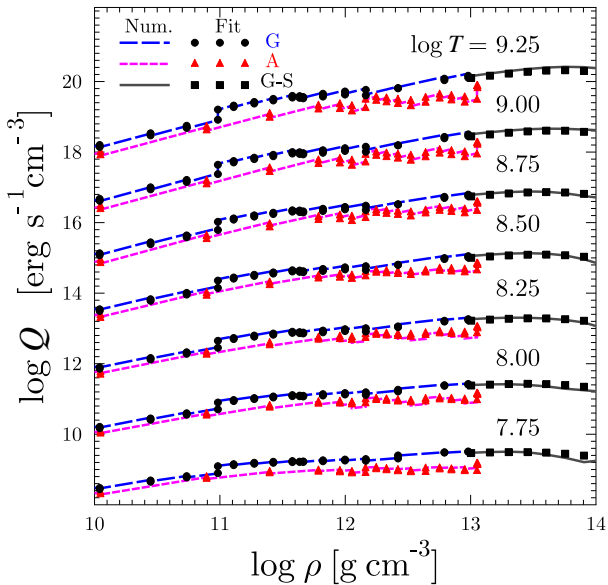}%{Q-rho_gro&acc.eps}
\caption{(Color online) Density dependence of neutrino
bremsstrahlung emissivity $Q$ for ground-state (G) and accreted (A)
crusts of neutron star at seven temperatures, $T=10^{7.75},~10^{8}$,
\ldots $10^{9.25}$ K. Lines are calculated using the formalism
 \cite{Kam1999}, while symbols are our fits. At $\rho \gtrsim
10^{13}$ g~cm$^{-3}$ the smooth composition model of ground-state
matter (G-S) is used (see text for details). }
\label{fig_Q-rho}
\end{figure}

After the optimization
the root-mean-square relative error of fitted $L$
becomes $8\%$, with the
maximum error $\sim 50\%$ at $Z=11$, $A=21$, $n_{\rm i} = 10^{34}$ cm$^{-3}$,
$T = 10^8$~K, and $R_{\rm p} = 0.022\,a$, in which case $L=0.059$.
This fit accuracy is quite sufficient
for applications.

By way of illustration, in fig.\ \ref{fig_logars} we show the
dependence of the full Coulomb logarithm $L$ (thick lines) and its
partial phonon contribution  $L_{\rm ph}$ (thin lines) versus
$\Gamma$ for $\tau=1$. We present three cases of $Z=6$ (solid lines)
and 26 (short-dashed lines) at $\xi=0$ (approximation of point-like
nuclei, typical for the outer crust) and $Z=50$ at $\xi=2$
(finite-size nuclei, characteristic for the bottom of the inner
crust). The lines are calculated using the formalism \cite{Kam1999}.
The kink of the solid line for $Z=6$ occurs at the melting point
because \cite{Kam1999} neglects the quantum effects of ion motion in
the liquid ($\Gamma < 175$). For the nuclei with larger $Z$ the
quantum effects in the liquid phase are markedly weaker and the kink
is smoothed out. Although formally $L_{\rm ph}$ is valid in the
crystal, we extend the curve in the liquid. Indeed, as already
mentioned before, $L_{\rm ph}$ reasonably well approximates the full
Coulomb logarithm in the solid and liquid phases excluding the
neighborhood of the melting point.

In the same fig.\  \ref{fig_logars} we plot our fits to the
full Coulomb logarithm $L$ (by dots, squares and triangles for
$Z=6$, 26 and 50, respectively).  The quality of the fits
is seen to be quite satisfactory.

Figure \ref{fig_Q-rho} displays the neutrino emissivitiy $Q$ as a
function of density $\rho$ in the neutron star crust at seven
temperatures, $T=10^{7.75}$, $10^{8}$, \ldots $10^{9.25}$ K. We use
two models of the crust, the ground-state and accreted one
\cite{Haens2007,Haens1990}. The ground-state crust corresponds to
the matter in thermodynamic equilibrium. The accreted crust is
formed by nuclear transformations in the matter compressed under the
weight of newly accreted material \cite{Haens1990}. The numerical
values of $Q$ for the ground-state crust are shown by the
long-dashed lines, and for the accreted crust by the short-dashed
lines. The fits are plotted by filled dots and triangles,
respectively. The accreted crust is composed of lighter nuclei with
lower $Z$. The bremsstrahlung neutrino emission in the accreted
crust is somewhat lower than in the ground-state one because of
lower $Z$ in the accreted crust. In both cases it is assumed that
only the nuclei of one type are present at any given density. Slight
jumps of the calculated and fitted $Q$ values at some densities are
associated with the change of nuclides in dense matter with growing
$\rho$ \cite{Haens2007}. Our fits reproduce numerically calculated
$Q$ values quite well.

At $\rho \gtrsim 10^{13}$ g~cm$^{-3}$ the accreted crust becomes
almost indistiguishable from the ground-state one, and we do not
plot the data for the accreted crust at higher densities. However,
the emissivity $Q$ for the dense ground-state crust starts to depend
on the proton density profiles within the nuclei (see above). To
demonstrate the quality of our fits in this regime we use the
smooth-composition model of the ground-state crust at $\rho \gtrsim
10^{13}$ g~cm$^{-3}$ (the solid lines), calculate the
root-mean-squared proton core radii and use them in our fits
(squares).  Again, there is good agreement of the theory and fits,
just as for kinetic properties of crustal matter \cite{Gned2001}.

%%%%%%%%%%%%%%%%%%%%%%%%%%%%%%%%%%%%%
\section{Effects of magnetic fields}
\label{s:mag}
%%%%%%%%%%%%%%%%%%%%%%%%%%%%%%%%%%%%

Many neutron stars possess strong magnetic fields which affect various
neutrino processes, including the process (\ref{bremsScheme}).
Accurate calculation of $Q$ for the process (\ref{bremsScheme}) is
a difficult and still unsolved problem. Here we
formulate the conditions at which magnetic fields can modify the
process. Evidently, the process can be affected by
magnetic fields through  electrons and atomic nuclei.

\begin{figure}
\onefigure[width=0.45\textwidth]{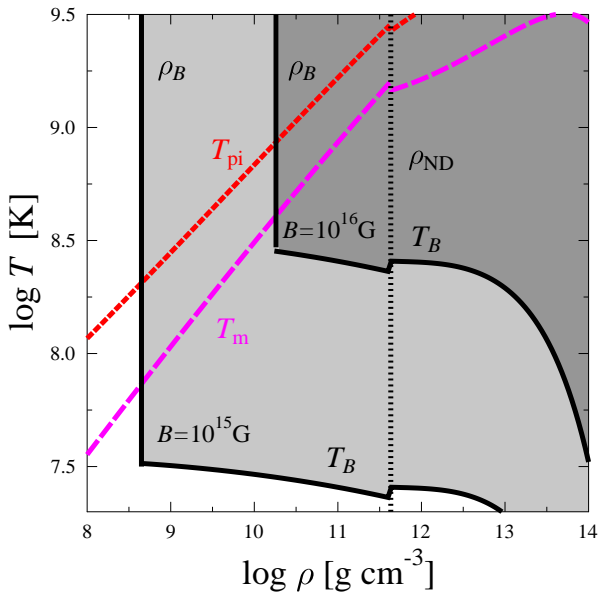}%{Q-rho_gro&acc.eps}
\caption{(Color online) Density-temperature diagram for the
ground-state neutron star crust. The densely shaded region is
expected to be almost unaffected by the magnetic field
$B=10^{16}$~G, while weakly shaded region is almost unaffected by
the field $B=10^{15}$~G. Any region is bounded by the density
$\rho_B$ below which the electrons occupy only the ground Landau
level, and by the temperature $T_B$ below which phonon frequencies
of Coulomb crystals of atomic nuclei are affected by $B$. Also shown
is the neutron drip density $\rho_{\rm ND}\approx 4.3 \times
10^{11}$ g~cm$^{-3}$, the melting temperature $T_{\rm m}$ of the
crystal and the ion plasma temperature $T_{\rm pi}$ (see text for
details). } \label{fig:phys}
\end{figure}

{\it Electrons.} A field $\bm{B}$ changes the motion of
electrons because of Landau
quantization of electron states; e.g. \cite{Haens2007}. In our
case of strongly degenerate relativistic electrons the importance of
magnetic effects is mostly determined by the characteristic density
\begin{equation}
\rho_B \approx 7045 (A_{\rm tot}/Z)B_{12}^{3/2}~{\rm g\ }
{\rm cm}^{-3},
\label{e:rhoB}
\end{equation}
where
$B_{12}=B/10^{12}$~G. At $\rho \lesssim \rho_B$ the electrons are
strongly quantized (occupy the ground Landau level)
while at $\rho \gg \rho_B$ they populate many Landau
levels.

Accordingly, we expect that the field $\bm{B}$ strongly affects the
process (\ref{bremsScheme}) at $\rho \lesssim \rho_B$. At higher
$\rho \gtrsim \rho_B$ the field affects the electron motion much
weaker. Such effects are regulated by the characteristic temperature
$T_{B\rm e}=\hbar \omega_{B\rm e}/k_{\rm B}$ associated with
gyrofrequency of degenerate electrons, $\omega_{B \rm e}=eB/(m_{\rm
e}c \sqrt{1+x^2})$. If $T \gtrsim T_{B\rm e}$ the thermal broadening
of the Landau levels exceeds the spacing between the levels and
washes out the Landau level structure. Then the magnetic fields can
be treated as non-quantizing; their effects on the process
(\ref{bremsScheme}) should be weak. At $\rho \gtrsim \rho_B$ but $T
\ll T_{B\rm e}$ the thermal broadening of Landau levels is smaller,
the structure of the Landau levels may be pronounced, and the
magnetic field behaves as weakly quantizing. It may produce not very
pronounced quantum oscillations of $Q$ with increasing $\rho$ due to
population of higher Landau levels. The emissivity $Q$ averaged over
oscillations is expected to resemble the field-free $Q$. Note that
Landau levels can also be broadened by electron collisions and other
delicate effects \cite{S84}, so that exact calculation of $Q$ can be
complicated. However, the magnetic effects of the electrons on $Q$
at $\rho \gtrsim \rho_B$ cannot be dramatic (can be neglected in the
first-order approximation).

{\it Atomic nuclei.} Strong magnetic fields affect vibration properties
(phonon modes) of Coulomb crystals of atomic nuclei and modify the neutrino-pair
bresmsstrahlung in this way. Phonon modes of magnetized Coulomb crystals have
been studied in a number of works, e.g.
%starting from the papers
%\cite{UGU80,NF82,NF83}. Detailed calculations of the
%phonon spectrum has been performed by
\cite{B09,BY13} and references therein. There are
three branches of phonon spectrum;
all of them can be affected
by magnetic fields. Various phonon modes are affected in certain
ranges of $\rho$ and $T$ (similar to those plotted in fig.\ 6 of \cite{BY13},
where magnetic fields modify phonon heat capacity). These effects are
mostly pronounced at
\begin{equation}
  T \ll T_{B}=\frac{\hbar \omega_{B}}{k_{\rm B}},\quad
    \omega_B=\frac{ZeB}{A_{\rm nuc}m_{\rm u}c},
\label{e:TBi}
\end{equation}
where $\omega_B$ is the ion cyclotron frequency. It is natural
to expect that at $T \gtrsim T_B$ the process (\ref{bremsScheme}) is
not affected by magnetic fields through the ion motion.

Figure \ref{fig:phys} shows possible ranges of $\rho$ and $T$ in the
neutron star crust where magnetic fields can influence the
neutrino-pair bremsstrahlung for the smooth-composition model of the
ground-state matter. We expect that the densely shaded region is not
affected by the magnetic field $B=10^{16}$~G, whereas the slightly
shaded region is not affected by the field $B=10^{15}$~G. Any of
these two regions is bounded by the density $\rho_B$, equation
(\ref{e:rhoB}),
 and by the temperature $T_B$,
equation (\ref{e:TBi}). We see that the fields $B \lesssim
10^{15}$~G have practically no effects on neutrino-pair
bremsstrahlung in the neutron star crust under formulated conditions
(see above), while higher fields may affect this neutrino process,
especially at low densities.

In addition, in fig.\ \ref{fig:phys} we plot the neutron drip
density $\rho_{\rm ND}\approx 4.3 \times 10^{11}$ g~cm$^{-3}$, the
melting temperature $T_{\rm m}$ of the crystal and the ion plasma
temperature $T_{\rm pi}$ for the ground state matter. All
these quantities are presented neglecting the
effects of magnetic fields.

%%%%%%%%%%%%%%%%%%%%%%%%%%%
\section{Conclusions}
%%%%%%%%%%%%%%%%%%%%%%%%%%

Using the formalism  \cite{Kam1999}, which includes rich spectrum of
physical effects,
we have derived a universal fit for the neutrino emissivity $Q$
of electron-nucleus bremsstrahlung (\ref{bremsScheme}) in a neutron star crust
(envelope). We have fitted
the expression for the generalized Coulomb logarithm $L$ which determines
$Q$. To this aim, we have used the method of effective potential $V_{\rm eff}$
of the electron-nucleus interaction and
proposed its form  (\ref{Lappr}).

We expect that our fit is valid for any realistic
composition of the neutron star crust, whereas the fit
presented in \cite{Kam1999} is obtained only for the ground-state
crust. Our fit is thought to be accurate in wide ranges
of parameters, particularly, in
the density range from about $10^8$ g~cm$^{-3}$ to the crust bottom
and in a wide temperature
range (from about a few times of $10^7$~K to a few times of
$10^9$~K). It is valid for the nuclei with
$6 \lesssim Z \lesssim 50$ and $1.9Z \lesssim A_{\rm nuc} \lesssim 3Z$.
The fit is convenient for using in
computer codes which simulate thermal evolution of neutron stars.
The fit assumes the presence of nuclei of one type
at any values of $\rho$ and $T$. In the case of multicomponent
plasma we recommend to use the approach of mean nucleus (mean ion), e.g.
\cite{Haens2007}.

As already mentioned before, the process (\ref{bremsScheme}) is the
leading neutrino process in a neutron star crust. It is important
for modelling transient phenomena in warm neutron stars. These
phenomena are associated with powerful processes of energy release
in the crust and/or thermal relaxation of the crust and core. In
particular, we mean thermal relaxation in young (age 10--100 yr)
isolated  neutron stars \cite{lrpp94,Gned2001}; thermal relaxation
in accreting neutron stars with overheated crust in soft X-ray
transients (after accretion stops and the crust equilibrates with
the core, as observed in MXB 1659--298, KS 1731--260 and some other
sources, e.g. \cite{degenaaretal13a,degenaaretal13b} and references
therein). The process (\ref{bremsScheme}) can also be important in
X-ray superbursts in accreting neutron stars (e.g. \cite{Gupta07}
and references therein), and in thermal evolution of magnetars in
quasistationary and bursting states \cite{m13}. In many cases the
composition of the crust is variable which affects the neutrino
emission and should be taken into account. In addition, there could
be isolated cooling neutron stars which are strongly superfluid
inside. Superfluidity suppresses the main neutrino processes in
stellar cores and makes neutrino emission from the crust important
for global thermal evolution \cite{ykg01}.

Our results can also be important for massive cooling white dwarfs
(or proto white dwarfs) or other stars with massive degenerate cores
(red giants and supergiants).

Although our results seem useful for many applications, the theory
of the process (\ref{bremsScheme}) needs to be elaborated. First, it
would be interesting to consider the process (\ref{bremsScheme}) in
a multicomponent plasma of atomic nuclei which would require
complicated calculations of corresponding structure factors. Second,
one needs to include the quantum effects of ion motion in the ion
liquid which is the long-standing but still unsolved problem. Third,
it would be important to study (\ref{bremsScheme}) for the electron
gas of any relativity and degeneracy; this can be done but requires
a lot of effort. Fourth, it would be interesting to analyze the
process (\ref{bremsScheme}) in possible phases of nuclear pasta of
highly non-spherical nuclear clusters which can exist in a layer
between the neutron star crust and the core. The latter problem
requires the Debye-Waller factors for such clusters \cite{Kam1999}
which are currently unavailable. Finally, it would be important to
take into account the effects of strong magnetic fields in neutron
stars. We have analysed the conditions at which very strong fields
can affect the process (\ref{bremsScheme}) but the influence of the
field on the process is still not explored and seems to be a
technically complicated task. The consideration of all these
problems goes far beyond the scope of this paper.

\acknowledgments
This work was supported by the Russian Science Foundation,
grant 14-12-00316.

\end{document}